\documentclass[twocolumn,showpacs,aps,prl,superscriptaddress,floatfix]{revtex4}

\usepackage{graphicx}
\usepackage{dcolumn}
\usepackage{bm}

\begin{document}

\preprint{}

\newcommand{\cro}{Cd$_2$Re$_2$O$_7$ } 

\affiliation{Department of Physics and Astronomy, McMaster University,
Hamilton, Ontario, L8S 4M1, Canada}
\affiliation{Solid State Division, Oak Ridge National Laboratory, P.O. Box 2008,
Oak Ridge, TN, 37831 U.S.A.} 
\affiliation{Department of Physics and Astronomy,
The University of Tennessee,Knoxville, TN, 37996 U.S.A.}
\affiliation{Canadian Institute for Advanced Research, 180 Dundas St. W., 
Toronto, Ontario, M5G 1Z8, Canada} 

\author{J.P. Castellan}
\affiliation{Department of Physics and Astronomy, McMaster University,
Hamilton, Ontario, L8S 4M1, Canada}
\author{B.D. Gaulin}
\affiliation{Department of Physics and Astronomy, McMaster University,
Hamilton, Ontario, L8S 4M1, Canada}
\affiliation{Canadian Institute for Advanced Research, 180 Dundas St. W., 
Toronto, Ontario, M5G 1Z8, Canada} 
\author{J. van Duijn}
\affiliation{Department of Physics and Astronomy, McMaster University,
Hamilton, Ontario, L8S 4M1, Canada}
\author{M.J. Lewis}
\affiliation{Department of Physics and Astronomy, McMaster University,
Hamilton, Ontario, L8S 4M1, Canada}
\author{M.D. Lumsden}
\affiliation{Solid State Division, Oak Ridge National Laboratory, P.O. Box 2008,
Oak Ridge, TN, 37831 U.S.A.} 
\author{R. Jin}
\affiliation{Solid State Division, Oak Ridge National Laboratory, P.O. Box 2008,
Oak Ridge, TN, 37831 U.S.A.} 
\author{J. He}
\affiliation{Department of Physics and Astronomy,
The University of Tennessee,Knoxville, TN, 37996 U.S.A.}
\author{S.E. Nagler}
\affiliation{Solid State Division, Oak Ridge National Laboratory, P.O. Box 2008,
Oak Ridge, TN, 37831 U.S.A.} 
\author{D. Mandrus}
\affiliation{Solid State Division, Oak Ridge National Laboratory, P.O. Box 2008,
Oak Ridge, TN, 37831 U.S.A.} 
\affiliation{Department of Physics and Astronomy,
The University of Tennessee,Knoxville, TN, 37996 U.S.A.}


\title{Structural Ordering and Symmetry Breaking in Cd$_2$Re$_2$O$_7$} 

\begin{abstract} 
Single crystal X-ray diffraction measurements have been carried out on 
\cro near and below the phase transition it exhibits at T$_{C'}\sim$195 
K.  \cro was recently
discovered as the first, and to date only, superconductor with the cubic
pyrochlore structure.  Superlattice Bragg peaks show an apparently
continuous structural transition at T$_{C'}$, however the order parameter 
displays anomalously
slow growth to $\sim$ $T_{C'}$/10, and resolution limited
critical-like scattering is seen above T$_{C'}$.  High resolution
measurements show the high temperature cubic Bragg peaks to split on
entering the low temperature phase, indicating a (likely tetragonal)
lowering of symmetry below T$_{C'}$. 
\end{abstract}

\maketitle 

Materials which crystallize into the cubic pyrochlore structure have been 
of intense recent interest, due to the presence of networks of 
corner-sharing tetrahedra within such structures \cite{reviews}.  Cubic 
pyrochlores 
display chemical composition A$_2$B$_2$O$_7$, and space group {\it 
Fd$\bar{3}$m}.  Independently, both the A and B 
sublattices reside on networks of corner-sharing 
tetrahedra, an architecture also common to Laves phase cubic spinels, for
example.  Such materials have the potential to display phenomena related 
to geometrical frustration in the presence of antiferromagnetism. 
While much activity has focused on local magnetic moments in insulating 
pyrochlores, interesting metallic properties have also been observed recently.
This has been the case, for example in Nd$_2$Mo$_2$O$_7$ \cite{Taguchi} 
where a large anomalous 
Hall effect has been measured, in Cd$_2$Os$_2$O$_7$ where a metal 
insulator transition occurs near 226 K \cite{CdOsO}, as well as 
in the spinel LiV$_2$O$_4$, which is the only known transition metal based 
heavy fermion conductor \cite{LVO}.

While many metallic, cubic pyrochlore oxides exist, no superconductors
were known to exist within this family of materials until very recently.  
Hanawa et al.\cite {Hanawa}, Sakai et al.\cite{Sakai}, and Jin et
al.\cite{Jin} have all recently reported superconductivity in \cro, the
first such pyrochlore. The superconducting T$_C$s are somewhat
sample dependent and have been reported between 1 and 2 K.  Moreover, the
relatively high temperature metallic properties are anomalous, and may be
driven by an as-yet poorly-understood phase transition near T$_{C'} \sim$
195 K \cite{Jin2}.  In this letter, we report on the nature of the phases
above and below T$_{C'}$ as well as the phase transition itself.  We show
compelling evidence for a splitting of the cubic Bragg peaks, indicating a
lowering of symmetry below T$_{C'}$, likely to a tetragonal structure, as
well as an unusual order parameter which grows very slowly with decreasing
temperature.

\cro is a rather poor metal near room temperature, exhibiting an almost
flat resistivity between 200 K and 400 K \cite{Hanawa,Sakai,Jin2}.  Just
below 200 K, the resistivity falls off sharply, continuing down to a low
temperature Fermi liquid regime characterized by a T$^2$ dependence
to the resistivity between 2 K and roughly 60 K, and a residual 
resistivity on the order of
10 $\mu$ ohm-cm \cite{Jin,Hanawa}.  On further lowering the temperature, 
the resistivity falls to zero, at T$_C$$\sim$ 1.4 K in crystals from the 
same
batch as that under study here, indicating the onset of the
superconducting state.  Heat capacity measurements show a large anomaly at
T$_C$, while above T$_C$ these measurements give a Sommerfeld $\gamma$
value of roughly 30 mJ/mol-K$^2$.  This result can be combined 
with A, the
coefficient of the quadratic term in the temperature dependence of the
resistivity, to give a Kadowaki-Woods ratio, A/$\gamma$, similar to that
seen in highly correlated metals such as the heavy fermion superconductor
UBe$_{13}$ \cite{Jin}. Recent transverse field $\mu$SR measurements 
\cite{Mark} reveal 
the presence of a vortex lattice below T$_C$, with a large and 
temperature independent value of the penetration depth below 0.4T$_C$.
These measurements show \cro  to be a type II superconductor and are 
consistent with a nodeless superconducting energy gap.

Heat capacity measurements show a pronounced anomaly near T$_{C'}$ $\sim$
200 K, consistent with a continuous phase transition\cite{Jin2}.  Electron
diffraction from single crystals\cite{Jin2} and preliminary x-ray 
diffraction studies
from powder samples\cite{Hanawa2} show the appearance of superlattice 
Bragg peaks at
reflections such as (0,0,6) and (0,0,10), which are inconsistent with the
(0,0,h):  h=4n condition appropriate to the high temperature cubic space
group.  In addition, DC susceptibility measurements\cite{Hanawa,Jin2} show 
an abrupt
reduction in the susceptibility below T$_{C'}$, similar to that seen in 
magnetic singlet ground state systems, such as CuGeO$_3$\cite{Hase} and 
NaV$_2$O$_5$\cite{Isobe}.

The high quality single crystal used in the present study was grown as
reported by He et al. \cite{He}.  It had approximate dimensions 4 $\times$ 
4 $\times$
2mm$^3$ and a mosaic spread of less than 0.04$^\circ$ full width at half 
maximum.  It was mounted in a Be can in the presence of a He
exchange gas.  The can was connected to the cold finger of a closed cycle
refrigerator with approximate temperature stability of 0.005 K near
T$_{C'}$, and 0.01 K elsewhere.  X-ray diffraction measurements were
performed in two modes, both employing an 18 kW rotating anode generator,
Cu K$\alpha$ radiation, and triple axis diffractometer.  Relatively low
resolution measurements of the superlattice Bragg peak intensities were
obtained using a pyrolitic graphite (002) monochromator and a scintillation
counter.  Much higher resolution measurements were performed with a
perfect Ge (111)  monochromator, and a Bruker Hi Star area detector,
mounted 0.75 m from the sample position on the scattered beam arm of
the diffractometer. The high resolution experiment easily separated
Cu$_{K\alpha_1}$ from Cu$_{K\alpha_2}$ diffraction, and was used for
precision measurements of the lineshapes of both the principal and
superlattice Bragg peaks.

Figure 1 shows the temperature dependence of the integrated intensity of
low resolution scans of the (0,0,10) superlattice Bragg peak.  Data was
taken in separate warming and cooling runs, and is shown over an extended
temperature range in the top panel, as well as over a narrow range near
T$_{C'}$ in the bottom panel.  The form of this scattering, proportional
to the square of the order parameter, is anomalous as it grows very slowly
below T$_{C'}$, especially between $\sim$ 170 K and $\sim$ 40 K.  A mean
field phase transition displays the slowest growth among conventional
models for cooperative behavior which exhibit continuous phase
transitions.  In Fig. 1, we compare the integrated intensity of the
superlattice Bragg scattering to the square of the mean field order
parameter\cite{Plischke}, as well as to the square of the order parameter 
appropriate to
the three dimensional Ising model\cite {3DIsing}.  Clearly the growth of 
the
measured order parameter is much slower with decreasing temperature, than
would be predicted by even mean field theory.

Closer to the phase transition, the measured order parameter does look 
more conventional as shown in the bottom panel of Fig. 1.  This panel 
shows 
integrated intensity with both downward curvature below $\sim$ 195 K, and 
upward curvature above $\sim$ 195 K, consistent with the measurement of 
the order parameter squared below T$_{C'}$, and the measurement of 
fluctuations in the order parameter, or so-called critical scattering, 
above T$_{C'}$.  Data in the temperature range 175 K to 190 K was analyzed
assuming this to be the case, and fits were carried out in the 
approximate range of reduced temperature from 0.02 to 0.10, placing the 
data within a typical asymptotic critical regime.  The fit of this data by 
the critical form:
\begin{equation}
Intensity=I_0 ( \frac{T_{C'}-T}{T_{C'}})^{2\beta}
\end{equation}
is shown in the bottom panel of Fig. 1.  The fit is of high quality, and 
it 
produces T$_{C'}$=194.3 $\pm$0.1 K and a critical exponent $\beta$=0.33 
$\pm$ 0.03, which is typical of three dimensional continuous phase 
transitions \cite{Collins}.

Although the (0,0,10) intensity above T$_{C'}$=194.3 appears to be 
critical scattering, we show below that it is anomolous and remains 
resolution limited at all temperatures measured.  {\bf Q}-broadened  
critical fluctuations can be measured by moving slightly off the (0,0,10) 
Bragg position, however it is extremely weak.  Scattering at 
(0.04,0.064,10) is shown in the bottom 
panel of Fig. 1 and it displays a weak peak near T$_{C'}$, as expected for a 
continuous phase transition.  This broad scattering is measured in 
counts per {\it hour}, and it makes a negligible contribution to the overall 
scattering around (0,0,10) above T$_{C'}$=194.3.  It does however provide 
a consistency check on the phase transition occuring at 194.3 K.

\begin{figure}[t] \centering
\includegraphics[width=0.95\columnwidth]{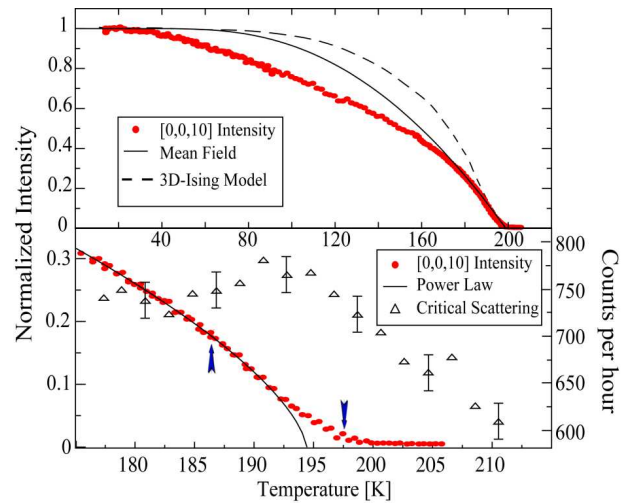} \caption{The 
top
panel shows the integrated intensity of the (0,0,10)  superlattice Bragg
peak as a function of temperature, compared with the square of the order
parameter expected from mean field theory, and that expected from the 3
dimensional Ising model.  The lower shows the same data in the immediate
vicinity of T$_{C'} \sim$ 194 K, along with a fit to critical behavior
modeled as a power law in reduced temperature (left-hand y-axis), and the
broad critical scattering as measured at (0.04,0.064,10) (right-hand 
y-axis).} \label{fig1}
\end{figure}

High resolution measurements of the lineshape of the (0,0,10) 
superlattice Bragg peak, shown in Fig. 2, reveal that almost all of the 
this scattering above T$_{C'}$=194.3 K remains resolution 
limited; it is indistinguishable from the superlattice 
scattering below 194.3 K, 
albeit weaker in intensity.  Figure 2 shows maps of the scattering at and 
around the (0,0,10) superlattice position at temperatures of 
185.7 K and 197.4 K.  As indicated by the arrows superposed on the order 
parameter shown in the bottom panel of Fig. 1, these temperatures 
correspond to well below and well above T$_{C'}$. 
 
\begin{figure}[t]
\centering
\includegraphics[width=0.95\columnwidth]{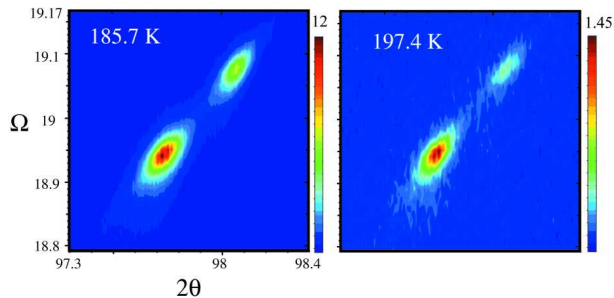}
\caption{High resolution scans of the scattering at and near the (0,0,10)
superlattice Bragg peak positions are shown for the temperatures marked by 
arrows in the inset to Fig. 1.  These temperatures correspond to well 
below and well above T$_{C'}$.
Data is shown in maps as a function of sample rotation angle $\Omega$ and 
scattering angle 2$\theta$.  The two diffraction features at 2$\theta$ 
values of $\sim$ 97.7$^\circ$ and 98.07$^\circ$ are from Cu 
K$_{\alpha1}$ and Cu K$_{\alpha2}$ radiation, respectively.  The top of 
the 
linear color scale is different for 
each data set.  It is clear that the lineshape does {\it not} broaden 
appreciably on passing through T$_{C'}$.}  
\label{fig2}
\end{figure}

The 197.4 K data set definitely falls within the upward
curvature regime of the temperature dependence of the superlattice 
integrated intensity, and would normally be expected to
be due to fluctuations in the order parameter rather than due to the order
parameter itself.  As discussed above, such critical scattering is 
expected to broaden in
{$\bf Q$}, and therefore in angular coordinates, as the temperature 
increases
beyond T$_{C'}$, indicating a finite and decreasing correlation
length.  Such broadening is clearly not observed in this scattering above 
T$_{C'}$. 

The anomalous nature of the lineshape of the superlattice scattering above
T$_{C'}$ may be due to ``second length scale''
scattering\cite{2lengthscale}, which has often been observed in high
resolution synchrotron x-ray experiments near phase transitions in
crystalline materials.  It is not fully understood, but has been
associated with the effect of near-surface quenched disorder on the phase
transition.  The x-rays in the present measurements have a penetration
depth of the order of ten microns (tens of thousands of unit cells)  and 
are thus not
particularly surface sensitive.  

It may also be that this superlattice
Bragg scattering is not the primary order parameter for the phase
transition near T$_{C'}$, but is a secondary feature, pulled along by 
the true underlying phase transition.  The anomolously slow growth of the 
(0,0,10) superlattice Bragg peak at low temperatures also supports such an
interpretation.
\begin{figure}[t] \centering
\includegraphics[width=0.95\columnwidth]{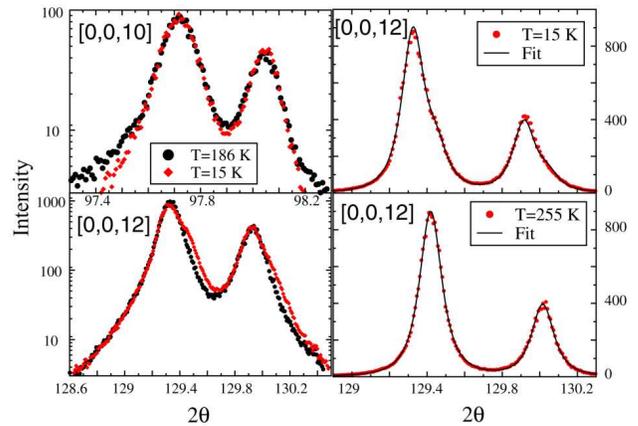}
\caption{The right hand panels shows longitudinal cuts taken through the
(0,0,12) principal Bragg peak position well above T$_{C'}$ (bottom) and
well below (top).  This data, on a linear scale, shows a clear shoulder on
the high 2$\theta$ side of both the Cu K$_{\alpha1}$ and K$_{\alpha2}$
peaks.  The left hand panels show the superlattice (0,0,10) (top) and
principal (0,0,12) (bottom) Bragg peaks on a semi-log scale, at
temperatures just below and well below T$_{C'}$.  Clearly, while the
principal Bragg peaks split on lowering the temperature, the superlattice
peaks do not.} \label{fig3} \end{figure}

We also carried out measurements of the principal Bragg peaks
(which satisfy the (00h): h=4n relation) at high scattering angle and 
in high resolution mode.  Scans along the
longitudinal direction, cutting through both Cu K$\alpha_1$ and Cu
K$\alpha_2$ peaks of the (0,0,12) principal Bragg peaks, are shown in the 
right hand panels of Fig. 3 on a linear scale, and in the bottom, left 
panel of Fig. 3 on a semi-log scale.  For comparison, similar longitudinal 
scans through the superlattice (0,0,10) Bragg peaks are shown on a semi-log 
scale in the top left panel of Fig. 3.  The data sets at the two 
temperature which make up both semi-log plots on the left side of Fig. 3 
have been shifted slightly in 2$\theta$ and, in the case of (0,0,10), 
scaled in intensity to allow for comparison.   

A shoulder is clearly seen to develop on the high angle side of the
(0,0,12) peaks as the temperature is lowered below T$_{C'}$.  This
indicates a splitting of the cubic Bragg peak into a lineshape
characterized by at least two different lattice parameters, and
consequently a symmetry lower than cubic, likely tetragonal, in the low
temperature state below T$_{C'} \sim$ 194.3 K.

A similar splitting of the (0,0,10) superlattice Bragg peak is {\it not} 
observed.  The superlattice Bragg peaks do not exist above T$_{C'}$ and 
thus we must compare superlattice data taken below, but near T$_{C'}$ with
that taken well below T$_{C'}$.  That is what is shown for both the 
principal (0,0,12) Bragg peak and the superlattice (0,0,10) Bragg peak at 
temperatures of 186 K and 15 K, respectively in the left hand panels of 
Fig. 3.  At 186 K, 
the splitting is not yet evident in either the principal, (0,0,12), or 
superlattice, (0,0,10),
Bragg peaks, but by 15 K it is clear in the principal Bragg peak, but not in 
the superlattice Bragg peaks. 
 
We fit the high temperature data at (0,0,12) to a phenomenological form at
255 K, assuming this to be the resolution-limited lineshape.  We then fit
(0,0,12) data at all temperatures to a form assuming the superposition of 
two such 
lineshapes, displaced
from each other in scattering angle (2$\theta$).  
This protocol allowed us to extract peak
positions for each of two peaks, giving the lattice parameters
as a function of temperature.  This analysis assumes that only two 
lattice parameters are
present at low temperatures, that is that the low temperature structure is
tetragonal.  The data at 15 K in the top right panel of Fig. 3 shows the
quality of this fit, and clearly the description of the data is
very good. The resulting tetragonal lattice parameters as a function of
temperature are shown in Fig. 4.

The behavior of the lattice parameters as a function of temperature is
striking.  The cubic lattice parameter displays the usual thermal
contraction with decreasing temperature until near T$_{C'}$.  The
splitting in the lattice parameters first develops near 200 K, with a 30 K
interval from $\sim$ 190 K to 160 K in which both lattice parameters
increase with decreasing temperature.  This trend continues to lower
temperatures for the larger of the two lattice parameters, while the 
smaller turns over below $\sim$ 150 K and displays relatively weak
contraction to lower temperatures.  At the lowest temperature measured, 15
K, the maximum splitting in lattice parameter measured is about 0.005~\AA~
or 0.05 $\%$.

The (0,0,10) superlattice peak does not show any splitting, and the 
temperature dependence associated with its periodicity is also shown in 
Fig. 4.  It clearly follows the upper branch of the two lattice parameters 
associated with (0,0,12).  These results imply that the low temperature 
state below T$_{C'}$ is likely twinned tetragonal, such that two of 
(0,0,12), (0,12,0) and (12,0,0) have one lattice parameter, while the 
other has a slightly different one.  The fact that the superlattice peak 
displays no splitting implies that only the subset of (0,0,10), (0,10,0) 
and (10,0,0) which follow the upper branch of the lattice parameter vs 
temperature curve shown in Fig. 4, exists.  The other(s) is (are) 
systematically absent.

\begin{figure}[t]
\centering
\includegraphics[width=0.95\columnwidth]{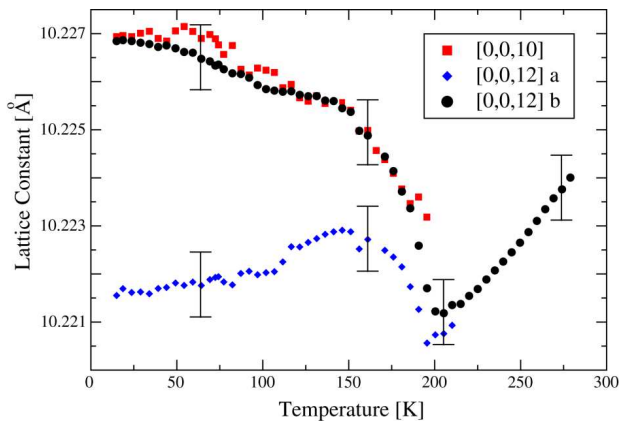}
\caption{The temperature dependence of the lattice parameters extracted 
from fits to the (0,0,12) principal Bragg peaks (shown as solid lines in 
Fig. 3) are shown, along the temperature dependence of the superlattice 
(0,0,10) Bragg peak periodicity below T$_{C'}$.}
\label{fig4}
\end{figure}

These observations allow us to discuss possible tetragonal space groups 
appropriate to the low temperature state below T$_{C'}$.  As the transition
appears to be close to continuous, we assume the low temperature state to 
be a gradual distortion of the high temperature cubic state, and thus the 
low temperature space group should be a subgroup of {\it Fd$\bar{3}$m}.  
There 
are two body-centered subgroups of {\it Fd$\bar{3}$m} which, in the presence 
of 
twinning, would split (0,0,12) and allow a single periodicity for (0,0,10).
These are {\it I4$_1$} and {\it I4$_1$22}.  In both cases the primitive 
unit 
vectors of 
the basal (a$^\prime$-a$^\prime$) plane are rotated by 45$^\circ$ relative 
to the 
unit 
vectors, a, of the high temperature cubic unit cell, and 
a$^\prime$=a/$\sqrt 2$.

While we cannot be more precise as to the low temperature space group at 
this time, we have unambiguously shown the symmetry of the low temperature 
state to be lower than cubic, and that this cubic symmetry breaking is 
an essential feature of the phase transition.  The most likely scenario is 
a cubic-tetragonal phase transition, with the primary order parameter being 
the difference in lattice parameters, given by $\sqrt 2$a$^\prime$-a. 
Finally we note that we do not expect this splitting of the 
lattice parameters to be evident in bulk measurements, such as dilatometry, 
unless a single domain sample can be produced at low temperatures.

We wish to acknowledge useful discussions with J.F. Britten.
This work was supported by NSERC of Canada.
Oak Ridge National Laboratory is managed by UT-Battelle, LLC 
the U.S. D.O.E. under Contract DE-AC05-00OR22725.  Work at UT is supported
by NSF DMR-0072998.


\begin{thebibliography}{}
\bibitem{reviews} 
for recent reviews see: S.T. Bramwell and M.J.P. Gingras, Science, 
{\bf 294}, 1495 (2001), {\bf Magnetic Systems with
Competing Interactions}, edited by H.T. Diep (World Scientific,
Singapore (1994).
\bibitem{Taguchi}
Y. Taguchi et al., Science, {\bf 291}, 2573 (2001).
\bibitem{CdOsO}
D. Mandrus et al., Phys. Rev. B, {\bf 63}, 195104 (2001).
\bibitem{LVO}
S. Kondo et al., Phys. Rev. Lett., {\bf 78}, 3729, (1997); C. Urano et 
al., Phys. Rev. Lett., {\bf 85}, 1052 (2000).
\bibitem{Hanawa}
M. Hanawa et al., Phys. Rev. Lett., {\bf 87}, 187001, (2001)
\bibitem{Sakai}
H. Sakai et al., J. Phys. Cond. Mat. {\bf 13}, L785 (2001).
\bibitem{Jin}
R. Jin et al., Phys. Rev. B, {\bf 64}, 180503(R), (2001)
\bibitem{Jin2}
R. Jin et al., cond-mat/0108402.
\bibitem{Mark}
M.D. Lumsden et al., cond-mat/0111187; R. Kadono et al., cond-mat/0112448.
\bibitem{Hanawa2}
M. Hanawa et al., cond-mat/0109050.
\bibitem{Hase}
M. Hase, I. Terasaki, and K. Uchinokura, Phys. Rev. Lett. {\bf 70}, 3651 
(1993).
\bibitem{Isobe}
M. Isobe and Y. Ueda, J. Phys. Soc. Jpn. {\bf 65}, 1178 (1996).
\bibitem{He}
J. He et al., unpublished.
\bibitem{Plischke} for example: M. Plischke and B. Birgerson, {\it
Equilibrium Statistical Physics, 2nd Edition}, World Scientific Press 
(1994).
\bibitem{3DIsing} D.M. Burley, Philos. Mag. {\bf 5}, 909 (1960).
\bibitem{Collins} M.F. Collins, {\it Magnetic Critical Scattering}, Oxford 
University Press (1989). 
\bibitem{2lengthscale} for example: S.R. Andrews, J. Phys. C {\bf 
19}, 3721 (1986).; T.R. Thurston et al, Phys. Rev. Lett. {\bf 70}, 3151 
(1993). 
\end{thebibliography}
\end{document}